\documentclass[twocolumn,showpacs,preprintnumbers,amsmath,amssymb]{revtex4}
\usepackage{graphicx}% Include figure files
\usepackage{dcolumn}% Align table columns on decimal point
\usepackage{bm}% bold math

\newcommand{\kbar}{$\bar{K}$}

\newcommand {\be}{\begin{equation}}
 \newcommand {\ee}{\end{equation}}
 \newcommand {\bea}{\begin{eqnarray}}
 
 \newcommand {\eea}{\end{eqnarray}}
 \newcommand {\kk}{$K^{-}K^{-}pp$}
 
\newcommand {\La}{\Lambda}

\begin{document}

\markboth{T. Yamazaki, Y. Akaishi and M. Hassanvand}{$K^-K^-pp$ Formation in $p+p$ collisions}

\title{
New Way to Produce Dense Double-Antikaonic Dibaryon System, $\bar{K}\bar{K} NN$,\\
 through $\Lambda(1405)$-Doorway Sticking in $p+p$ Collisions}

%\subtitle{This is a Subtitle}

\author{Toshimitsu{\sc Yamazaki}$^{1,2}$, Yoshinori {\sc Akaishi}$^{1,3}$, Maryam {\sc Hassanvand}$^{1,4}$ }

\address{$^{1}$ RIKEN Nishina Center, Wako, Saitama 351-0198, Japan}
\address{$^{2}$ Department of Physics, University of Tokyo, Tokyo 113-0033, Japan}
\address{$^{3}$ College of Science and Technology, Nihon University, Funabashi, Chiba 274-8501, Japan}
\address{$^{4}$ Department of Physics, Isfahan University of Technology. Isfahan 84156-83111, Iran
}

%\thanks{} 
\date{Proc. Jpn Acad, Ser. B {\bf 87} (2011); Received Jan. 6, 2011, Accepted Apr. 19, 2011}

\begin{abstract}
A recent successful observation of a dense and deeply bound $\bar{K}$ nuclear system, $K^-pp$, in the $p + p \rightarrow K^+ + K^-pp$ reaction in a DISTO experiment indicates that the double-$\bar{K}$ dibaryon, $K^-K^-pp$, which was predicted to be a dense nuclear system, can also be formed in $p+p$ collisions. We find theoretically that the $K^-$-$K^-$ repulsion plays no significant role in reducing the density and binding energy of $K^-K^-pp$ and that,  when two $\Lambda(1405)$ resonances are produced simultaneously in a short-range $p+p$ collision, they act as doorways to copious formation of $K^-K^-pp$, if and only if $K^-K^-pp$ is a dense object, as predicted.
\end{abstract}

\maketitle

\section{Introduction}

For the past decade we have predicted and studied deeply bound and dense kaonic nuclear states using an empirically based coupled-channel $\bar{K}N$ complex potential \cite{Akaishi02,Yamazaki02,Dote04a,Dote04b,Yamazaki04,Yamazaki07a,Yamazaki07b}. The structure of the most basic system, $K^-pp$, first predicted in 2002 \cite{Yamazaki02}, was studied in detail by a realistic three-body calculation \cite{Yamazaki07a,Yamazaki07b}, and hence a molecular nature of the strong binding was revealed. The $K^-pp$ system was shown to be close to a $\Lambda^*$-$p$, where $\Lambda^* \equiv \Lambda(1405)$ is a quasi-bound $I=0$ $\bar{K}N$ pair, like an ``atom". This study led us to a new concept of nuclear force, ``super-strong nuclear force", which is caused by a migrating real $\bar{K}$ between two nucleons as in molecular covalency \cite{Heitler}. It has a binding strength nearly 4-times as large as that of the ordinary nuclear force. This ``kaonic origin of nuclear force" is contrasted to the ordinary ``pionic origin of nuclear force" by Nishijima \cite{Nishijima08}.

\begin{figure}[htb]
\vspace{-0.5cm}
\includegraphics[width=0.45\textwidth]{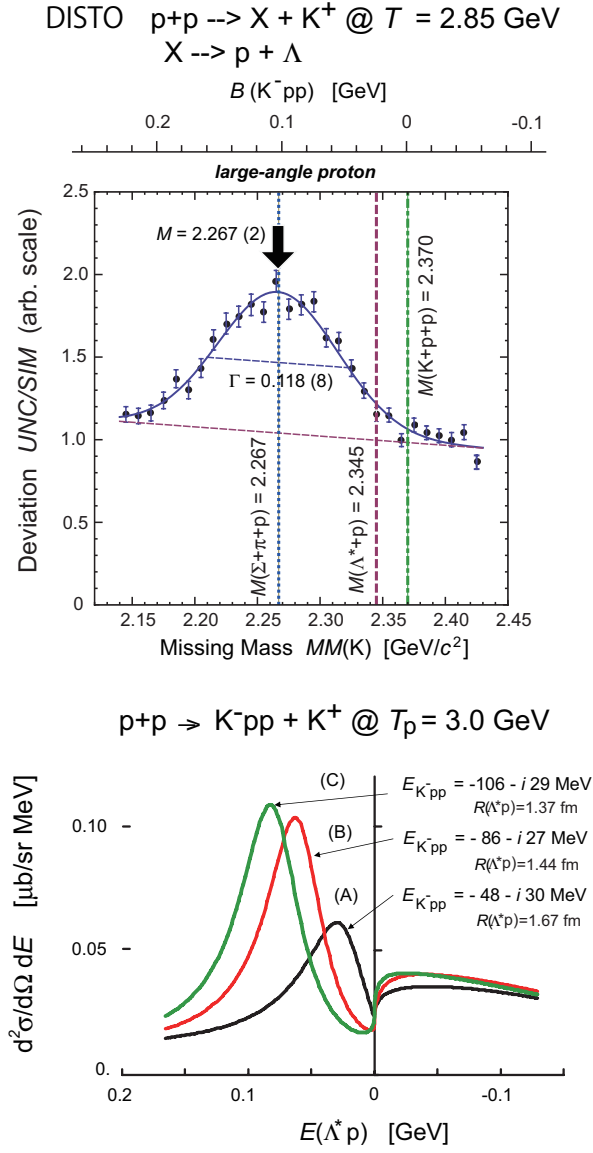}%
\caption{\label{fig:DISTO} (Upper) Observed $DEVIATION$ spectrum of the missing-mass of $K^+$, showing a dominant peak of $K^-pp$, in $pp \rightarrow K^+ + K^-pp$ \cite{Yamazaki10}. (Lower) Theoretical mass (energy) spectra of $K^-pp$ in the same reaction for three versions of the $\bar{K}N$ interaction \cite{Yamazaki07b}. The experimental data seem to be compatible with the version C, namely, the 25\% enhanced one. }
\end{figure}

Since the $p+p$ collision at high energy is known to produce $\Lambda^*$ among other hyperons, as revealed in $\Delta M(p K^+)$ missing-mass spectra \cite{MM,Zychor}, we proposed a nuclear reaction to populate and identify $K^-pp$ \cite{Yamazaki07b},
\begin{eqnarray} \label{eq:pp2KpL*}
p + p &\rightarrow& K^+ + \underline{p + \Lambda^*},\\
                                                 && ~~~~~~~~~~~ \hookrightarrow p \Lambda^* \rightarrow K^-pp, \nonumber
\end{eqnarray} 
with a subsequent decay:
\begin{equation}
          K^-pp \rightarrow  p + \Lambda.
\end{equation}
The observed production cross section of $\Lambda^*$ at an incident proton energy, $T_p \sim 3$ GeV, is about 10\% of the cross section, $\sigma _{\Lambda} \sim 50~ \mu$b, for the ordinary $\Lambda(1115)$. It was theoretically clarified that the produced $\Lambda^*$ serves as a doorway to form a complex $\Lambda^*$-$p \approx K^-pp$. This sticking process occurs strongly when the formed $K^-pp$ has a small $p$-$p$ distance that matches the small proximity of the $p+p$ collision (with a short collision length of $\sim \hbar /m_{\rho} c  \sim 0.3$ fm helped by a large momentum transfer of $\sim$ 1.6 GeV/$c$ at $T_p \sim 3$ GeV). In other words, the doorway particle, $\Lambda(1405)$, sticks to the participating $p$ at a high probability to form a $K^-pp$ state, if and only if it is dense, and thus the occurrence of this reaction would provide definite evidence for the dense $\bar{K}$ nuclear state, as predicted. 
  
Very recently, this reaction process was searched for in old DISTO experimental data of  exclusive events of $pp \rightarrow p + \Lambda + K^+$ at $T_p = 2.85$ GeV/$c$, and was indeed found to take place \cite{Yamazaki10}. A broad peak showing $M = 2267 \pm 2 (stat) \pm 5 (syst)$ MeV/$c^2$ was revealed with as much intensity as free $\Lambda^*$ emission, as shown in Fig.~\ref{fig:DISTO} (Upper). Thus, this experiment has proven that the state observed is a dense $K^-pp$. Additional evidence that the state is strongly related to $\Lambda^*$ production in the entrance channel has been obtained from the incident energy dependence of the cross section \cite{Kienle11}. The observed mass of $K^-pp$ in DISTO is much smaller than the original prediction of \cite{Yamazaki04,Yamazaki07b}. Two more versions of predictions in which the $\bar{K}N$ interaction is enhanced by 17\% (B) and 25\% (C) are given in \cite{Yamazaki07b}, as reproduced in the same figure. The version (C), 25\% enhanced one, seems to be compatible with the observation. An independent indication for a similar state in $K^-$ absorption by light nuclei was reported from a FINUDA experiment \cite{FINUDA}. There are also theoretical calculations of $K^-pp$ by the Faddeev method \cite{Shevchenko07a,Ikeda07,Shevchenko07b}.

It is extremely interesting and important to extend the search to double-$\bar{K}$ nuclear clusters, since they may serve as precursors to kaon condensation \cite{Nelson,Brown,Brown-Bethe}. Previously in 2004, we predicted the simplest double-$\bar{K}$ nuclear clusters, $K^-K^-pp$ and $K^-K^-ppn$, using the original AY interaction combined with the $G$-matrix method \cite{Yamazaki04}, which were shown to have deeper binding and higher density than $K^-pp$. Invariant-mass spectroscopy for their decay particles, $\Lambda + \Lambda$ and $\Lambda + \Lambda + p$, was proposed therein. In view of the DISTO data, the $\bar{K}N$ interaction and thus the actual binding of $K^-K^-pp$ may be much stronger than in the original prediction. How can one produce such double $\bar{K}$ nuclei? This is a crucial and important question. There is an exotic proposal to make use of antiproton absorption in $^4$He \cite{Kienle07,Zmeskal09}.  

In the present short communication we propose a new method to form $\bar{K}\bar{K}NN$ nuclear clusters in $p+p$ collisions and point out some important aspects. This was triggered by the successful observation of $K^-pp$ in the $p+p$ collision. We extended our thinking about the case of $pp \rightarrow K^+ + K^-pp$ to a new process, $pp \rightarrow K^+ + K^+ + K^-K^-pp$, and carried out its formulation and numerical calculations. A detailed report on the results and implications will be given elsewhere \cite{Hassanvand11}. Let us first discuss the structure of $K^-K^-pp$.

\section{Structure of $K^-K^-pp$}

\begin{figure}[htb]
\vspace{-0.5cm}
\includegraphics[width=0.4\textwidth]{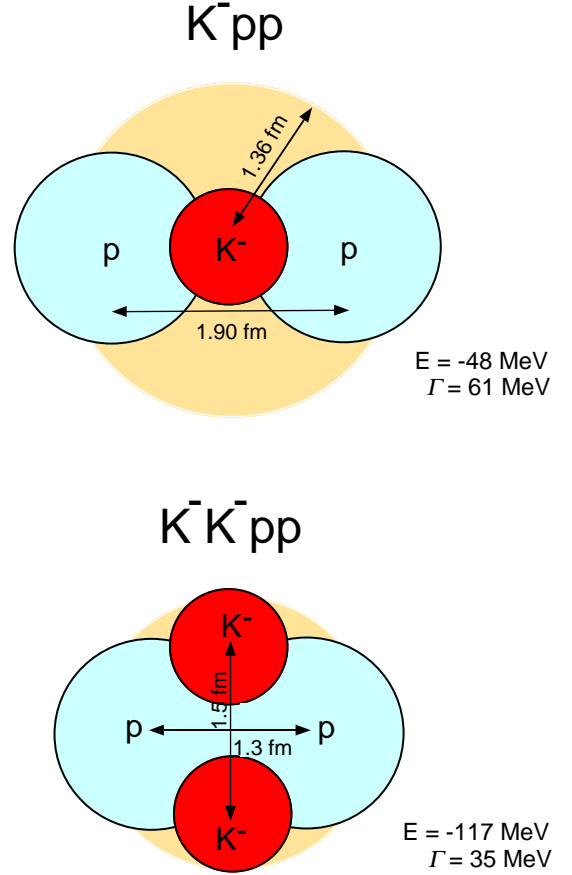}%
\caption{\label{fig:ppKK-dango}Schematic structure diagrams for the calculated $K^-pp$ and $K^-K^-pp$ nuclei. The rms radius of $K^-$ and the rms inter-nucleon and inter-\kbar~distances are shown. }
\end{figure}

Schematic structures of calculated $K^-pp$ and $K^-K^-pp$ states are shown in Fig.~\ref{fig:ppKK-dango}. The former was comprehensively studied in \cite{Yamazaki07b}, where a variational method (ATMS) \cite{Akaishi86} was employed. The structure is well approximated by a molecule type, with $\Lambda^* = K^-p$ as a ``hydrogen atom", which is coupled with another proton, and the $K^-$ meson migrates between the two protons, causing a super-strong nuclear force. This mechanism is understood to be a general mechanism for producing dense nuclear states around $\bar{K}$. 

The structure of $K^-K^-pp$ was calculated by using the same AY interaction, as reported in \cite{Yamazaki04}. The binding energy is $B_{KK} = 117$ MeV, much larger than that in $K^-pp$, $B_K = 48$ MeV. The mean distance of $p-p$ was found to be 1.3 fm for $K^-K^-pp$, whereas it is 1.9 fm for $K^-pp$. Namely, $K^-K^-pp$ is deeper bound and denser than $K^-pp$. As in the case of $K^-pp$ (see Fig. 4 of \cite{Yamazaki07b}), $K^-K^-pp$ is approximately composed of $\Lambda^*$-$\Lambda^*$, where the two $K^-$ mesons migrate between the two protons. 
Thus, we take the following interaction form between $\Lambda^*$-$\Lambda^*$ to represent  arbitrarily chosen level energy and width of $K^-K^-pp$:    
\begin{equation}
H_{\La^*\La^*} = -\frac{\hbar^2}{2\mu_{\La^*\La^*}}\vec{\nabla}^2 + U_{\La^*\La^*}(r),
\label{eq:Hamiltonian}
\end{equation}
where
\begin{equation}
U_{\La^*\La^*}(r) =  (V_0+iW_0) \, (\frac{r}{b})^2 \, {\rm exp}(-\frac{r}{b}) 
\label{eq:U}
\end{equation}
with potential strength parameters, $V_0$ and $W_0$, and a range $b$, which is related to the rms distance of $\Lambda^*$-$\Lambda^*$. 

\begin{figure}[htb]
\vspace{0.5cm}
\includegraphics[width=0.48\textwidth]{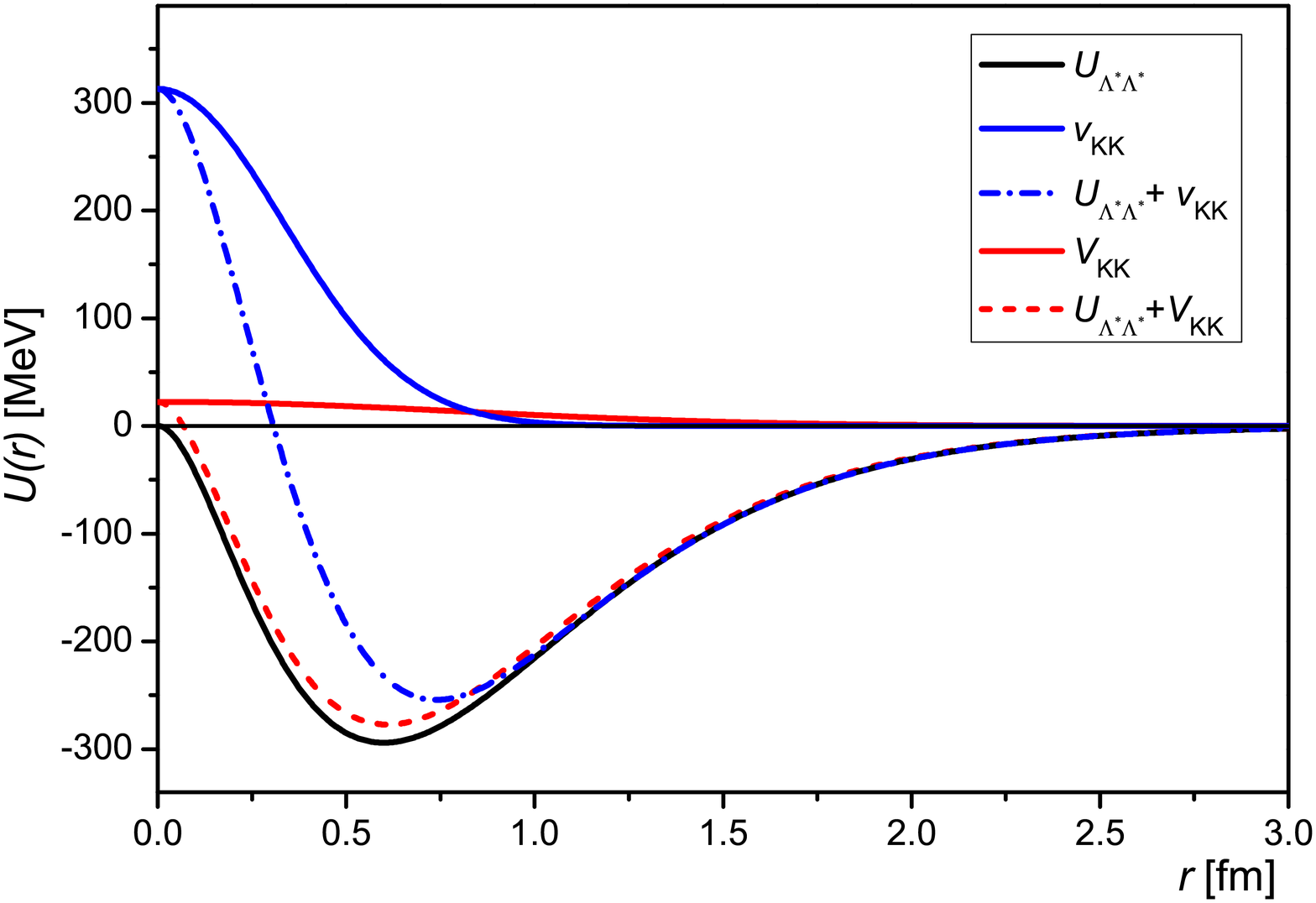}
\caption{\label{fig:KK} (Black solid curve) The $\Lambda^*$-$\Lambda^*$ potential for the case of $E = -150 - i \, 75$ MeV. (Blue solid and dash-dotted curves) The bare $K^-$-$K^-$ potential without finite-size correction, $v_{KK}$, and the incorporated $\Lambda^*$-$\Lambda^*$ potential, respectively. (Red solid and dashed curves) The effective $K^-$-$K^-$ potential with finite-size correction, $V_{KK}$, and the incorporated $\Lambda^*$-$\Lambda^*$ potential, respectively. }
\end{figure}

Since the experimentally identified $K^-pp$ has a substantially deeper binding ($B_K \sim 105$ MeV as compared with the originally predicted value of 48 MeV) \cite{Yamazaki10}, corresponding to a 25\% enhanced $\bar{K}N$ interaction, $K^-K^-pp$ is likely to have a much larger binding energy than the originally predicted one. We thus presume that the binding energy can be 150 MeV or more. 
For such cases we use the $\Lambda^*-\Lambda^*$ model Hamiltonian to describe any arbitrarily chosen energy $E$ in the following formalism and calculation. The $E$ we use is defined with respect to the emission threshold of $\Lambda^* + \Lambda^*$ and is thus connected to $B_{KK}$ and $M$ as $B_{KK} = -(E - 2 \times 27 ~{\rm MeV})$ and $M = (2 M_p + 2 m_K ) +  (E - 2 \times 27 ~{\rm MeV})/c^2$.
For the time being we take the bound $K^-K^-pp$ energy to be
\begin{equation}
E = -150 - i\, 75~{\rm MeV}
\end{equation}
as a standard one. This energy is produced by the $\Lambda^*-\Lambda^*$ potential with $V_0 = -543$ MeV and $W_0 = -181$ MeV for $b = 0.3$ fm. As shown in Fig.~\ref{fig:KK} (black solid curve), this potential form expresses molecule-type bonding like the Morse potential, characterized by a short-range repulsion together with a long-range attraction. 

So far, no extra interaction between the two $K^-$ mesons has yet been taken into account. We now consider the $K^-$-$K^-$ interaction, which is constructed by Kanada-En'yo and Jido \cite{Kanada} to be: 
\begin{equation}
v_{\bar{K} \bar{K}}(r_{KK}) = v_0 \, {\rm exp}\, [-(\frac{r_{KK}}{b_{KK}})^2],
\end{equation}  
with $v_0 = 313$ MeV and $b_{KK} = 0.47$ fm, based on the result of a lattice QCD calculation \cite{KK}. This looks a very repulsive interaction, as shown in Fig.~\ref{fig:KK} (blue solid curve), which would reduce the binding of the two $\Lambda^*$'s significantly. If we could add the above interaction to $H_{\La^*\La^*}$ (blue dashed curve), it would bring a reduction of the $K^-K^-pp$ binding energy by 34 MeV. However, this procedure is not valid; when we incorporate the above interaction into $H_{\La^*\La^*}$, we have to correct for the finite size of $\Lambda^*$'s, in which the $K^-$ mesons are orbiting. We now take the $K^-$ meson distribution in $\Lambda^*$ to be expressed by
\begin{equation}
\rho_K (r_K) = (\frac{a^3}{\pi^{3/2}} ) \, {\rm exp}[- (\frac{r_K}{a})^2]
\end{equation}
with $a = 1.11$ fm, corresponding to an rms distance of $R_{rms} = 1.36$ fm \cite{Yamazaki07b}. 
%The density distribution of $K^-$ from the c.m. of $\Lambda^*$ is written as
%\begin{equation}
%\rho_K (r_K) = (\frac{1}{\pi^{3/2}} \tilde{a}^3) \, {\rm exp}(- (\frac{r}{\tilde{a}})^2),
%\end{equation}
%with $\tilde{a} = M_p/ (M_p + m_K) a$. 
Then, after the correction for the center of mass, we obtain a double folding potential for the finite size:
\begin{equation}
V_{KK} (r) = v_0 \frac{1}{f^3} \, {\rm exp} [- (\frac{r}{f \, b_{KK}} )^2],
\end{equation}
with
\begin{equation}
f = \sqrt{1 + 2 ( \frac{a} {b_{KK}} \frac{M_p}{M_p + m_K} )^2 }\sim 2.41.
\end{equation}
Effectively, the inclusion of the finite size of $\Lambda^*$ enlarges the range parameter, $b_{KK}$, by a factor of $f = 2.4$, 
and weakens the repulsive potential strength by a factor of $f^{-3} \sim 0.072$. For a point-like $K^-$ distribution, $a \rightarrow 0$ and $f \rightarrow 1$. Thus, the folded $K^-$-$K^-$ interaction is lowered and longer-ranged, as shown in Fig.~\ref{fig:KK} (red solid curve), and the total $\Lambda^*$-$\Lambda^*$ potential is slightly increased, as shown in Fig.~\ref{fig:KK} (red dashed curve). The $K^-$-$K^-$ interaction plays no significant role in the structure and formation of $K^-K^-pp$; it yields a reduction of the binding energy of $K^-K^-pp$ by only $\sim 13$ MeV. 

It is often said that kaon condensation is unlikely because of the repulsion among $\bar{K}$'s. When the kaon condensed matter is composed of $\Lambda^*$ particles through the super-strong nuclear force, as envisaged in Ref.\cite{Yamazaki07b}, then, the bare $K^-$-$K^-$ repulsion is suppressed greatly from the above consideration of the finite size of $\Lambda^*$. 

 \begin{figure}
    %Requires \usepackage{graphicx}
    \center\includegraphics[width=7cm]{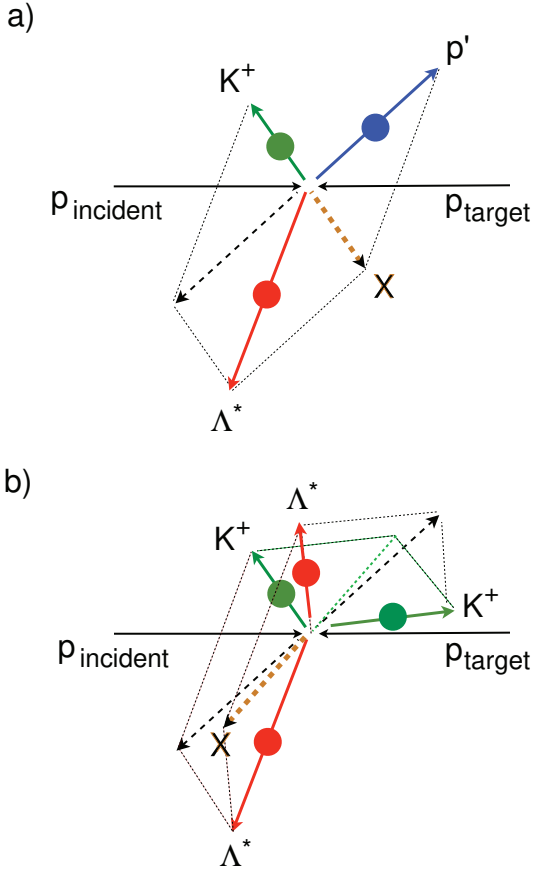}
    \caption{\label{fig:pp2KKL*L*} Reaction diagrams involving $\Lambda^* = \Lambda(1405)$ as a doorway in the center-of-mass system of $p+p$ collisions. a) For $pp \rightarrow p + K^+ + \Lambda^* \rightarrow K^+ + K^-pp$ and b) for $pp \rightarrow K^+ +  \Lambda^* + K^+ +\Lambda^* \rightarrow K^+ + K^+ + K^-K^-pp$.  In a) the vector sum of the $\Lambda^*$ and $K^+$ momenta is balanced by the scattered-proton momentum, and in b) this is replaced by another pair of $\Lambda^*$ and $K^+$. The momentum of $X$, expressed by a sepia bold broken line, is composed of those of $p$ and $\Lambda^*$ in a) and those of $\Lambda^*$ + $\Lambda^*$ in b). }
    \end{figure}

\section{$K^-K^-pp$ formation in the $p+p$ collision}

The ingredients of the nuclear reactions that we consider are shown in Fig.~\ref{fig:pp2KKL*L*}. Case a) is for the formation of $K^-pp$ \cite{Yamazaki07b}, in which a $\Lambda^*$ is produced with a large momentum transfer at nearly the same position as the participating proton within a very short collision length of $\hbar /m_{\rho} c \sim$ 0.3 fm. This $\Lambda^*$ sticks to the participating scattered proton to form $X = K^-pp$, as indicated by a sepia bold broken line. The vector sum of the $\Lambda^*$ and $K^+$ momenta is balanced by the momentum of the scattered proton. Likewise, as shown in b), the $p+p$ collision at a high enough energy produces two $\Lambda^*$'s together with two $K^+$'s. Here, the proton in Case a) is replaced by another pair of $\Lambda^*$ and $K^+$, and the two $\Lambda^*$'s stick to each other, forming $X = K^-K^-pp$, as shown.

The free $\Lambda$ production cross section in $p+p$ collision is known to be
\begin{equation}
\sigma_{\Lambda} \sim 10^{-3} \times \sigma_{total} \sim 50~\mu {\rm b},
\end{equation} 
and the free $\Lambda^*$ production cross section at 2.83 GeV is known from Ref.\cite{Zychor} to be
\begin{equation} \label{eq:sigmaL1405}
\sigma _{\Lambda^*} \sim 4.5~\mu {\rm b} \sim 0.10 \times  \sigma _{\Lambda} ,
\end{equation} 
No free double-$\Lambda^*$ production is known at all, and thus, only a rough estimate can be made here. The production of two normal $\Lambda$'s is expected to take place at a rate of
\begin{equation} 
\sigma _{\Lambda+ \Lambda}  \sim 0.001 \times \sigma_{\Lambda} \sim 50~{\rm nb}.
\end{equation} 
Thus, the free double-$\Lambda^*$ production can be 
\begin{equation} \label{eq:sigmaL1405L1405}
\sigma _{\Lambda^* + \Lambda^*} \sim 0.01 \times \sigma _{\Lambda + \Lambda} \sim 0.5~{\rm nb}.
\end{equation} 
Since the production of a single $\Lambda^*$, eq.(\ref{eq:pp2KpL*}), is associated with a short-range collision ($m_B \sim m_{\rho}$), its cross section is small, but the same hard collision can produce another $\Lambda^*$. So, the cross section for double $\Lambda^*$'s may not be so small as given in eq.(\ref{eq:sigmaL1405L1405}).  
These very rough estimates, of course, depend on the incident energy. We lack good experimental and theoretical  information. The problem here, however, is to investigate how much fraction of $\sigma _{\Lambda^* + \Lambda^*}$ can contribute to the formation of the bound $K^-K^-pp.$
 
Let us make an extension of the reaction (\ref{eq:pp2KpL*}) to double-$\bar{K}$ production:
\begin{eqnarray} \label{pp2L1405L1405}
p &+&p \nonumber\\  
&\rightarrow& K^+ + K^+ + \underline{\Lambda^* + \Lambda^*},\\
                                                 && ~~~~~~~~~~~~~~~~~~~ \hookrightarrow \Lambda^*\Lambda^* \rightarrow K^-K^-pp,\nonumber 
\end{eqnarray}
with a subsequent decay:
\begin{equation}
K^-K^-pp \rightarrow \Lambda + \Lambda, 
\end{equation}
and others.

\begin{figure}
%\Requires \usepackage{figures}
\center\includegraphics[width=8cm]{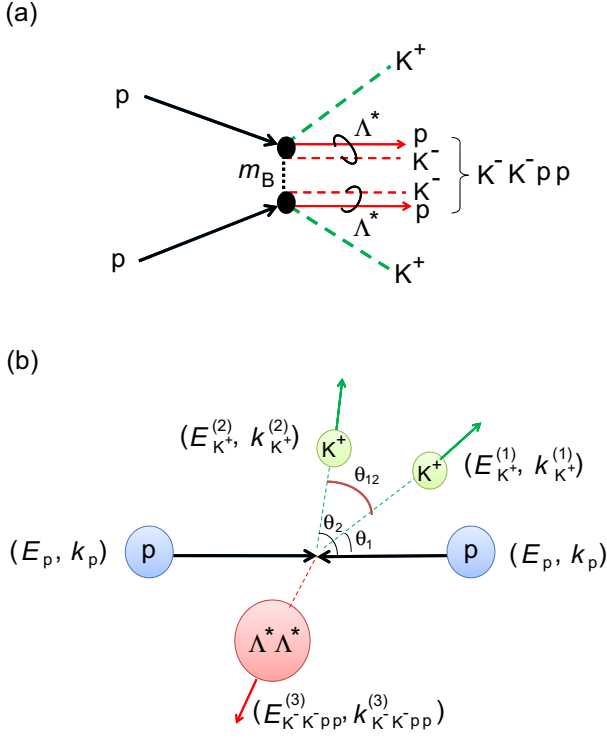}\\
\caption{(a) Elementary process for $p + p \rightarrow K^+ +K^++K^-K^-pp$ via the production of two $\Lambda^*$'s as doorways. The intermediate boson with a mass $m_B$ is shown. (b) Kinematical event pattern for the production of two $K^+$'s and a $\Lambda^* \Lambda^*$.} \label{fig:mechanism}
\end{figure}

Then, the same mechanism as proven experimentally in the case of $K^-pp$ can be applied to $K^-K^-pp$ formation, where the simultaneous production of two doorway $\Lambda^*$'s within a short distance, given by the intermediate boson with a mass $m_B$, helps their sticking into a dense $\Lambda^*\Lambda^*  \approx K^-K^-pp$. 
Thus, we have formulated and calculated the cross sections according to Fig.~\ref{fig:mechanism}, where two $K^+$'s with $(E^{(1)}, \vec{k}^{(1)})$ and $(E^{(2)}, \vec{k}^{(2)})$, and a $\Lambda^* \Lambda^*$ complex with $(E^{(3)}, \vec{k}^{(3)})$ are shown. A detailed report will be given elsewhere \cite{Hassanvand11}. The essential ingredients are as follows.\\

 1) A large momentum transfer $Q = |\vec{k_p} - \vec{k^{(3)}}| \sim 1.8$ GeV/$c$ in the $p-p$ reaction at $T_p \sim 7$ GeV.\\
 
  2) A short collision range of $\beta \sim \hbar/m_{\rho} c \sim$ 0.26 fm.\\
  
  3) Compactness of $K^-K^-pp$ (a short distance of $\Lambda^*$-$\Lambda^*$). \\

\begin{figure}
\center\includegraphics[width=8.5cm]{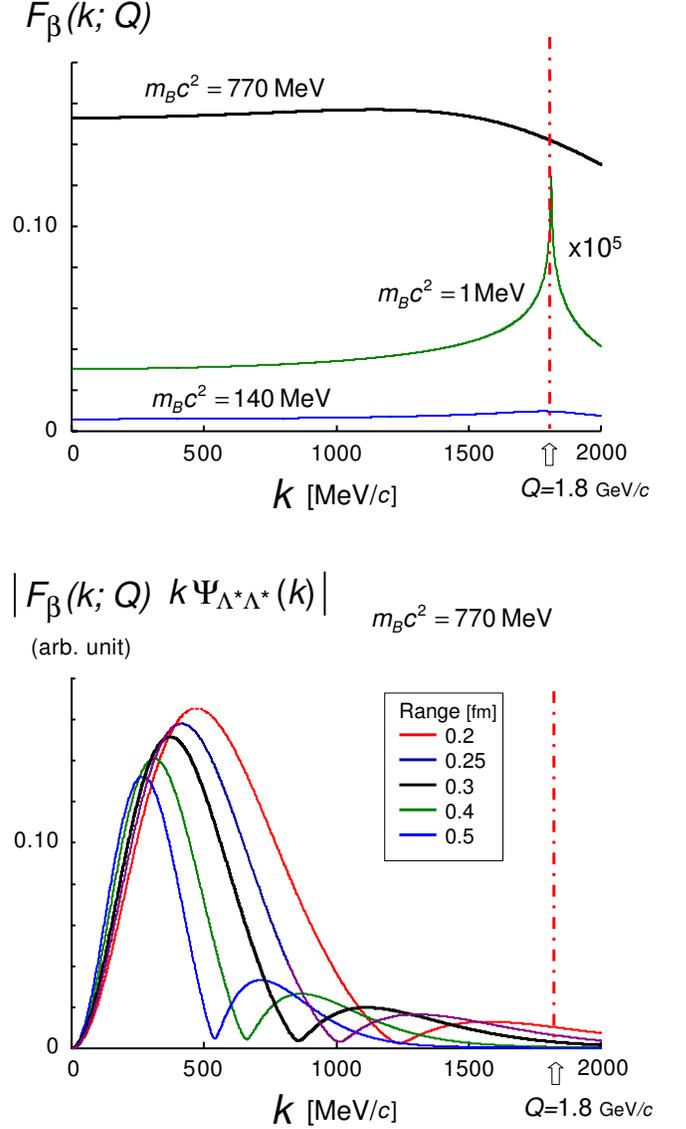}\\
\caption{(Upper) The momentum window function $F_{\beta} (k; Q)$.  (Lower) The integrands of the formation amplitude as a function of the relative momentum of $\Lambda^*$-$\Lambda^*$, which are nearly the same as the wavefunctions $|\Psi (k)|$, for various values of $b$.}\label{fig:A}
\end{figure}

  These three variables are physically independent, but collaborate jointly in the formation of a compact $K^-K^-pp$ in high-energy $p+p$ collisions.  
   The elementary process, 7 GeV $p \rightarrow K^+ + \Lambda^*$, alone cannot occur in the free space, since the momentum conservation is not fulfilled. This process is realized only when some part of the momentum is transferred to another $\Lambda^*$ by virtual meson exchange with a mass $m_B$, which we express in terms of the interaction, 
\begin{equation}
\frac{{\rm exp} (-r/\beta)} {\beta^2\, r}
\end{equation}
with
\begin{equation}
\beta = \frac{\hbar}{m_B c}.
\end{equation} 
Therefore, this interaction is inevitable in the present $\Lambda^* \Lambda^*$ formation reaction. If two $\Lambda^*$'s move together as a result of the momentum transfer, the $\Lambda^*$-$\Lambda^*$ relative momentum becomes much smaller than $Q=1.8$ GeV$/c$ and may match the Fourier transform:
\begin{equation}
\frac{\Psi_{\Lambda^* \Lambda^*} (k)}{k}Y_{00}(\hat k) = \frac{1}{(2\pi)^3} \int d \vec r~{\rm exp}(-i \vec k \vec r)~ \phi_{\Lambda^* \Lambda^*}(\vec r),
\end{equation} 
for the bound-state wavefunction $\phi_{\Lambda^* \Lambda^*}(\vec r)$. The bound-state wavefunction distributes  centered around 400 MeV/$c$, which justifies our use of the non-relativistic Schr\"odinger equation. 
So, we investigate the relative-momentum distribution which contributes to the formation amplitude ($A$) of the reaction:
\begin{eqnarray}
A &=& \int d \vec r~{\rm exp}(-i \vec Q \vec r) \, \frac{{\rm exp}(-r/\beta)}{\beta^2 r}\, \phi_{\Lambda^* \Lambda^*}(\vec r)~~~~~~ \\
 &=& (4\pi)^{5/2} \int_0^\infty dk ~F_\beta (k;Q) ~k \Psi_{\Lambda^* \Lambda^*}(k),
\end{eqnarray} 
where 
\begin{eqnarray}
 F_\beta (k~;Q) &=& \frac{1}{(4\pi)^2} \int d \hat k \nonumber \\
 &\times& \int d \vec r~{\rm exp}(-i (\vec Q+\vec k) \vec r)~ \frac{{\rm exp}(-r/\beta)}{\beta^2 r} \nonumber\\
&=&\frac{1}{4Qk\beta^2} \ln \frac{1+\beta^2(Q+k)^2}{1+\beta^2(Q-k)^2}.
\end{eqnarray}

 The function $F_{\beta} (k; Q)$ expresses a momentum window function, as shown in Fig.\ref{fig:A} (upper). It approaches $\rightarrow 1$ in the limit of $\beta \rightarrow 0$, and its scale decreases as $\beta$ increases. The window function with $m_B = m_{\pi}$ ($\beta = 1.41$ fm) is much smaller than that with  $m_B = m_{\rho}$ ($\beta = 0.26$ fm). 
Figure \ref{fig:A} (lower) shows the integrand of the formation amplitude as a function of the relative momentum ($k$) between the two $\Lambda^*$'s. Due to the effect of the interaction, exp$(-r/\beta)/(\beta^2r)$ with $m_B=770$ MeV/$c^2$, the function distributes in a rather flat form below the momentum transfer $Q=1.8$ GeV$/c$. Then, the intergrand functions of $A$ resemble the relative-momentum distributions of the bound-state wavefunction $\Psi(k)$, which range from 200 to 1000 MeV/$c$, depending on the range ($b$) of $\Lambda^*$-$\Lambda^*$.

In many cases of ordinary reactions the momentum window, $F$, is given as  
\begin{eqnarray}
F(k~;Q) &=& \frac{1}{(4\pi)^2} \int d \hat k \int d \vec r~{\rm exp}[-i (\vec Q+\vec k) \vec r)]\, \frac{1}{L^3 } \nonumber \\
&=& \frac{\pi}{2L^3Q^2} \delta (k-Q) 
\end{eqnarray}
with a normalization length ($L$). In this case the integrand of $A$ concentrates at $k=Q$ and an accurate description of high-momentum component of the wavefunction is required, but this is not the case of our present reaction. 

\begin{figure}
  \center\includegraphics[width=8.5cm]{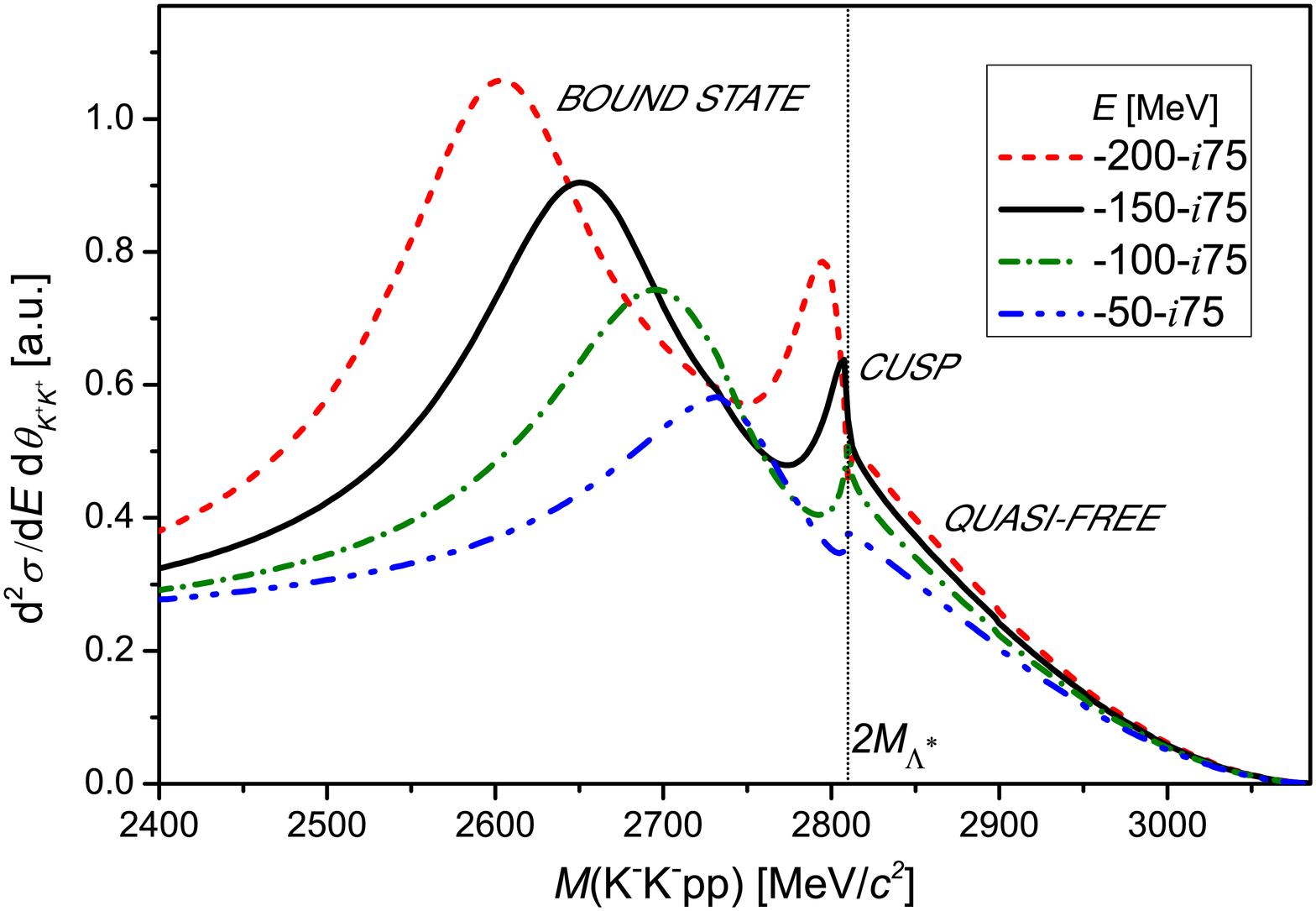}\\
    \caption{ Differential cross sections for various bound-state energies of the $K^-K^-pp$ system, $E$. For ${T_p}$ = $7.0$ GeV, $b$ = $0.3$ fm and $\theta_{12}$ = $180^{\circ}$. }\label{fig:DCS-Strength}
    \end{figure}

The production cross section of {\kk} ($\approx {\La^*\La^*}$) at energy $E$ is given, besides the factor composed of the relevant kinematical variables, by a dynamical spectral function:
\begin{eqnarray}
&&S(E)=-\frac{1}{\pi}  \nonumber \\
&&~\times~ {\rm Im} \big[ \int\!\int d\vec{r'}d\vec{r}f^*(\vec{r'})\langle\vec{r'}|\frac{1}{E-H_{\La^*\La^*}+{i} \epsilon}|\vec{r}\rangle f(\vec{r})\big] \nonumber\\
&&=-\frac{8\mu_{\La^*\La^*}}{\hbar^2}\sum_{\ell=0}^\infty(2\ell+1){\rm Im} \big[ \frac{1}{W(u_\ell^{(0)}u_\ell^{(+)})} \nonumber\\
&&\times \int_{0}^{\infty} d\vec{r'}\int_{0}^{\infty} d\vec{r} \, {\rm exp}(  -\frac{r'}{\beta})j_\ell(Q r')u_\ell^{(0)}(r_{<}) \nonumber\\ 
&&\times u_\ell^{(+)}(r_{>})j_\ell(Q r){\rm exp}(  -\frac{r'}{\beta}) \big], \label{eq:spectral-fn}
\end{eqnarray}
with 
\be
f(\vec{r})=\frac{\beta}{r}{\rm exp}(  -\frac{r}{\beta}+i\vec{Q}\vec{r}),
\ee
where $u_\ell^{(0)}(r)$ and $u_\ell^{(+)}(r)$ are the stationary and outgoing solutions of the Schr\"{o}dinger equation, respectively, and $W$ is their Wronskian. The ${\La^*\La^*}$ Hamiltonian is given by (\ref{eq:Hamiltonian},\ref{eq:U}).

%%%%%%%%%%%%%%%%%%%5
%%%%%%%%%%%%%%%%%%%%

We calculated the differential cross sections under various conditions.  Figure \ref{fig:DCS-Strength} shows the calculated cross sections at an incident proton energy of 7 GeV. The level energy of $K^-K^-pp$ with respect to the $\Lambda^* + \Lambda^*$ emission threshold is assumed to be $E$ = -200, -150, -100 and -50 MeV with a width of 150 MeV; also, the range of the $\Lambda^*$-$\Lambda^*$ interaction in the form (\ref{eq:U})
 is assumed to be $b = 0.3$ fm, which corresponds to $R_{\rm rms} = 0.94$ fm for the $E = -150$ MeV case.

\begin{figure}
\center\includegraphics[width=8.5cm]{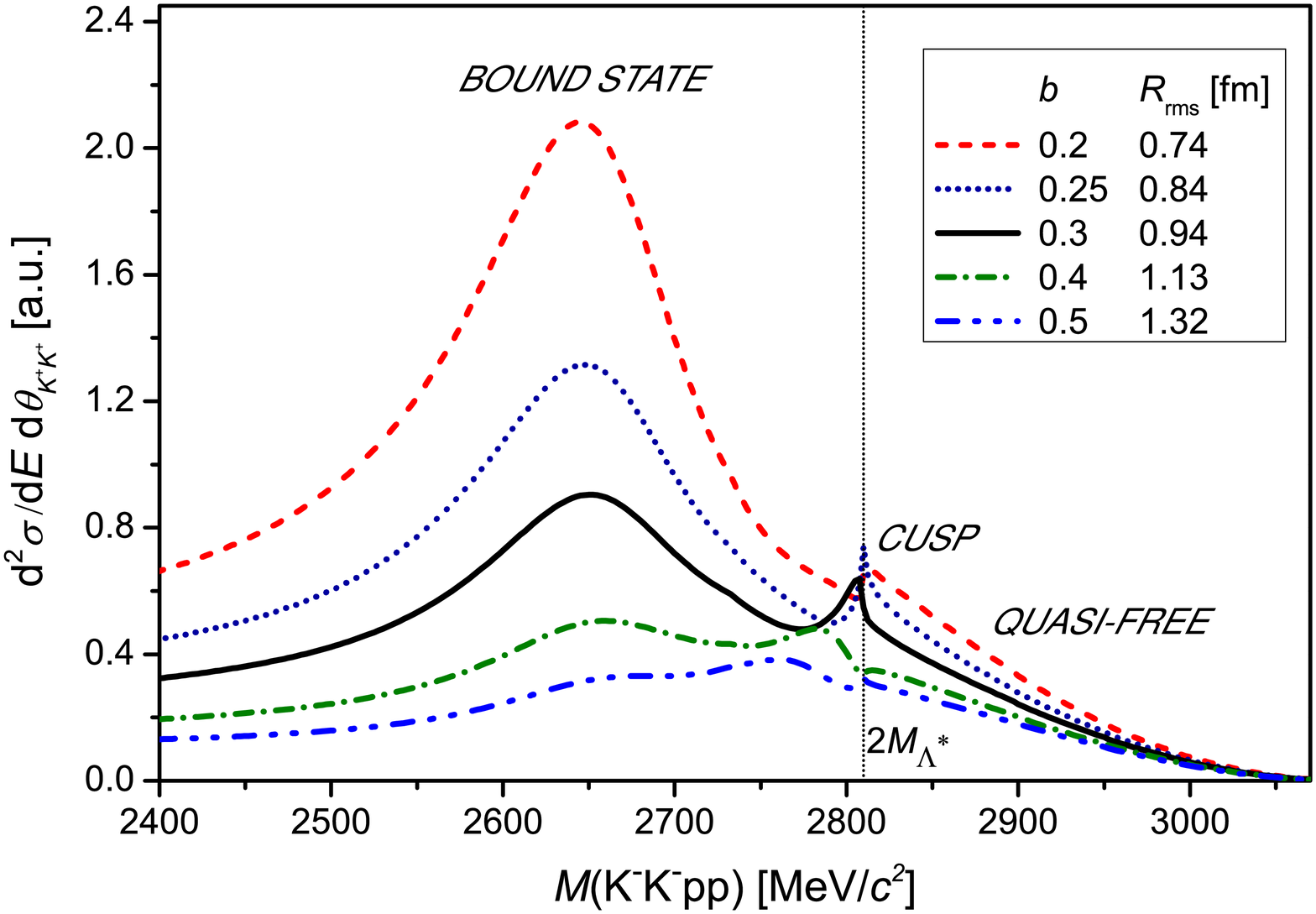}\\
\caption{ Differential cross sections of $p + p \rightarrow K^+ + K^+ + K^-K^-pp$ with $E = -150 - i\, 75$ MeV for various range parameters, $b$, which correspond to different $\Lambda^*$-$\Lambda^*$ rms distances, $R_{rms}(\La^*\La^*)$, as shown in the inset. For ${T_p}$ = $7.0$ GeV and $\theta_{12}$ = $180^{\circ}$.}\label{fig:DCS-rms}
\end{figure}

Each spectrum shows a characteristic shape composed of three parts:
\begin{itemize}
\item {\bf QF} (quasi-free): $M > M(2 \Lambda^*)$,
\item {\bf CUSP} (cusp): $M \approx M(2 \Lambda^*)$,
\item {\bf BS} (bound-state): $M \approx M(K^-K^-pp)$.
\end{itemize}

The continuum QF component extends from $M(2 \Lambda^*)$ to the kinematical limit. This might be supposed to be the main part, dominating the whole spectrum. Surprisingly, however, in the bound region $M < M(2 \Lambda^*)$, a large peak corresponding to $M(K^-K^-pp)$ dominates the spectrum. This situation is quite similar to the case of $p+p \rightarrow K^+ + K^-pp$ formation, and reflects the compact nature of $K^-K^-pp$. Near the threshold mass, $M(2 \Lambda^*)$, we predict a narrow structure, which can be explained as being a threshold cusp or an excited bound state. Such a narrow structure is smeared out by the large width of $\Lambda^*$. 
From this figure we recognize that the BS structure is enhanced as the binding energy (and thus the density) increases, in accordance with intuition. We find the BS/QF ratio to be
$BS/QF = 0.43, 1.09$, and 1.21 
for $E = -100, -150$, and -200 MeV, respectively. 
 The peak intensity dominates over the QF continuum region, indicating that the $K^-K^-pp$ production occurs with the same probability as the free emission of two $\Lambda^*$'s.  

To see the situation more clearly, we set $E$ to the standard case, -150 MeV, and varied the range, $b$, for which we obtained different potential parameters. Figure~\ref{fig:DCS-rms} shows the differential cross sections for various ranges of the interaction. The range parameters we varied, $b = 0.2, 0.25, 0.3, 0.4$ and 0.5 fm, correspond to different $rms$ distances of $\Lambda^*$-$\Lambda^*$    as $R_{\rm rms} = 0.74, 0.84, 0.94, 1.13$ and 1.32 fm, respectively. This figure demonstrates clearly that a more compact system is more favorably populated with a larger cross section. The reason for this is well explained in terms of the relative momentum $k$ of $\Lambda^*$-$\Lambda^*$ in Fig.~\ref{fig:A}. The BS/QF ratio is seen to be 2.51, 1.60, 1.09 and 0.36 for $b =$ 0.2, 0.25, 0.3 and 0.4 fm, respectively. This behavior can be used to determine $R_{\rm rms}$ from an observed spectrum of $M(K^-K^-pp)$. In other words, an observation of BS will indicate that $R_{\rm rms} <$ 1.0 fm.

\begin{figure}[h]
\centering
  \includegraphics[width=8.5cm]{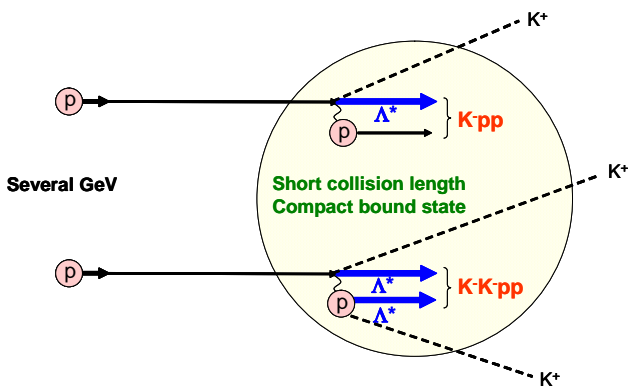}
  \caption{Schematic picture of the formation of basic kaonic nuclear clusters, $K^-pp$ and $K^-K^-pp$, which are doorways to {\it Swan Nuclear Physics}.} \label{fig:Swan}
\end{figure}

\section{Concluding remarks: toward "SWAN Nuclear Physics"}

Stimulated by a recent successful observation of the most basic single-$\bar{K}$ nuclear cluster, $K^-pp$, in the reaction $p + p \rightarrow K^+ + K^-pp$ \cite{Yamazaki10} in agreement with the proposed reaction mechanism of "enhanced sticking of $\Lambda^*$-$p$ doorway" \cite{Yamazaki07b}, we have presented a further idea that the simplest double-$\bar{K}$ nuclear cluster, $K^-K^-pp$, can be produced efficiently in the reaction $p + p \rightarrow K^+ + K^+ + K^-K^-pp$, and found that a similar kind of "$\Lambda^*$-$\Lambda^*$ enhanced-sticking" mechanism exists; the $K^-K^-pp$ cluster can be produced as much as the free production of $\Lambda^*$ + $\Lambda^*$ with the proton energy around 7 GeV, if it is a deeply bound and dense object. Detailed accounts will be published elsewhere  \cite{Hassanvand11}. Such a proton beam will be available in J-PARC in Japan and also in the FAIR project in Darmstadt. Since $K^-K^-pp$ is the most important gateway toward kaon condensation physics, we hope experimentalists will pursue how to realize such an experiment. 

We have also found that the $K^-$-$K^-$ repulsion inside $K^-K^-pp$ gives only a small effect on its structure, which does not alter the dense nature of double-$\bar{K}$ clusters. In $\bar{K}$ nuclei the $s$ quark combined with $\bar u$ or $\bar d$ quark in $\bar{K}$ plays a unique and dominant role in forming dense nuclear systems. The essence of physics here is {\it i)} the strong attraction of intruder $(\bar{u}, \bar{d})$ quarks with surrounding $(u, d)$ quarks without short-range repulsive barrier, and in addition, {\it ii)} the super-strong nuclear force caused by migrating $\bar{K}$ mesons, as clarified in \cite{Yamazaki07a}. We call this new paradigm {\it Swan Nuclear Physics}, since an alternative naming of {\it SWAN} given for the strangeness (and the second generation of elementary particles) by the late Professor K. Nishijima \cite{Nishijima:06} is quite suitable for expressing the essentially important role of $s$ quarks. High-energy $p+p$ collisions populate effectively the kaonic nuclear clusters, $K^-pp$ and $K^-K^-pp$, by satisfying two conditions of $a)$ short collision length and $b)$ compactness of produced clusters. Thus, several-GeV $p+p$ collisions in a target nucleus provide just efficient doorways to {\it Swan Nuclear Physics}, as depicted in Fig.~\ref{fig:Swan}. Needless to say, this paradigm is deeply connected to the physics of kaon condensation and $\bar{K}$-condensed matter and stars \cite{Nelson,Brown,Brown-Bethe}, and eventually to the physics of cold and dense quark matter.  

Finally, we note that the present reaction mechanism (high-momentum transfer high-sticking process for the formation of a dense kaonic cluster) provides an efficient way to {\it directly} forming $K^-pp$ and $K^-K^-pp$ clusters, not only in the $p+p$ reaction but also in $p$-nucleus and heavy-ion reactions, in contrast to the thermal equilibrium processes \cite{BM}. For the direct production processes, large incident energies, such as 3 GeV for $K^-pp$ and 7 GeV for $K^-K^-pp$, are necessary. Thus, high-energy reactions with kinetic energies of more than 7 GeV/N will be an interesting field to produce multi-$\bar{K}$ nuclear systems \cite{Yamazaki04}.  \\

%\section{Doorways to "Swan Nuclear Physics"}

 We are grateful to Professor P. Kienle for stimulating discussion. One of us (M.H.) would like to thank Professor S. Z. Kalantari for helping her before coming to Japan. This work is supported by 
Grants-in-Aid for Scientific Research of Monbu-kagaku-sho of Japan and of Ministry of Science, Research and Technology of Iran.


\begin{thebibliography}{99}


\bibitem{Akaishi02} Akaishi, Y. and Yamazaki, T. (2002)
                    Nuclear $\bar{K}$ bound states in light nuclei. 
                    Phys. Rev. C {\bf 65}, 044005. % 1-9 
\bibitem{Yamazaki02} Yamazaki, T. and Akaishi, Y. (2002) 
                    ($K^-,\pi^-$) production of nuclear $\bar{K}$ 
                    bound states in proton-rich systems via ${\Lambda}^*$ 
                    doorways. Phys. Lett. B {\bf 535}, 70-76.
\bibitem{Dote04a} Dot$\acute{\rm e}$, A., Horiuchi, H., Akaishi, Y. 
                    and Yamazaki, T. (2004) 
                    High-density $\bar{K}$ nuclear systems with isovector 
                    deformation. Phys. Lett. B {\bf 590}, 51-56. 
\bibitem{Dote04b} Dot$\acute{\rm e}$, A., Horiuchi, H., Akaishi, Y. 
                 and Yamazaki, T. (2004) 
                    Kaonic nuclei studied based on a new framework of 
                    antisymmetrized molecular dynamics. 
                    Phys. Rev. C {\bf 70}, 044313. % 1-11 
\bibitem{Yamazaki04} Yamazaki, T., Dot$\acute{\rm e}$, A. and Akaishi, Y. 
                  (2004) Invariant-mass spectroscopy for condensed single- and 
                    double-$\bar{K}$ nuclear clusters to be formed as residues 
                    in relativistic heavy-ion collisions. Phys. Lett. B 
                    {\bf 587}, 167-174.
\bibitem{Yamazaki07a} Yamazaki, T. and Akaishi, Y. (2007) 
                    Super strong nuclear force caused by migrating $\bar{K}$ 
                    mesons - Revival of the Heitler-London-Heisenberg scheme 
                    in kaonic nuclear clusters. 
                    Proc. Jpn. Acad., Ser. B {\bf 83}, 144-150.
\bibitem{Yamazaki07b} Yamazaki, T. and Akaishi, Y. (2007) 
                    Basic $\bar{K}$ nuclear cluster, $K^-pp$, and its enhanced 
                    formation in the $p + p \rightarrow K^+ + X$ reaction.
                    Phys. Rev. C {\bf 76}, 045201. % 1-16
\bibitem{Heitler} Heitler, W. and London, F. (1927) 
                    Wechselwirkung neutraler Atome und homoopolare Bindung 
                    nach der Quantenmechanik.
                    Z. Phys. {\bf 44}, 455-472. 
\bibitem{Nishijima08} Nishijima, K. (2008) cover page and caption of Proc. Jpn. 
                    Acad., Ser. B {\bf 84}, issue 7.
\bibitem{MM} Maggiora, M. {\it et al.} (2001) New results from DISTO for spin 
                    observables in exclusive hyperon production. 
                    Nucl. Phys. A {\bf 691}, 329c-335c.
\bibitem{Zychor} Zychor, I. {\it et al.} (2008) 
                    Lineshape of the $\Lambda$(1405) hyperon measured through 
                   its $\Sigma^0\pi^0$ decay.
                    Phys. Lett. B {\bf 660}, 167-171.
\bibitem{Yamazaki10} Yamazaki, T. {\it et al.} (2010) 
                    Indication of a deeply bound and compact $K^-pp$ state 
                    formed in the $pp \rightarrow p \Lambda K^+$ reaction at 2.85 GeV. Phys. Rev. Lett. {\bf 104}, 132502.
\bibitem{Kienle11} Kienle, P. {\it et al.} (2011) 
		Role of $\Lambda(1405)$ in the Formation of $X = K^-pp$ Revealed in $pp \rightarrow X + K^+$ at 2.50 		and 2.85 GeV. Phys. Lett. B (submitted).
\bibitem{FINUDA} Agnello, M.  {\it et al.} (2005)
                    Evidence for a kaon-bound state $K^-pp$ produced in $K^-$ 
                    absorption reactions at rest.
                    Phys. Rev. Lett. {\bf 94}, 212303. 
%\bibitem{Faber:09} M. Fabar, A.N. Ivanov, P. Kienle, J. Marton and M. Pitschmann, J. Mod. Phys. E (December 27th, 2010).
\bibitem{Shevchenko07a} Shevchenko, N.V., Gal, A. and Mares, J. (2007) 
                    Faddeev calculation of a $K^-pp$ quasibound state.
                    Phys. Rev. Lett. {\bf 98}, 082301.
\bibitem{Ikeda07} Ikeda, Y. and Sato, T. (2007) 
                    Strange dibaryon resonance in the $\bar{K}NN-\pi YN$ system.
                    Phy. Rev. {\bf C 76},  035203.
\bibitem{Shevchenko07b} Shevchenko, N.V., Gal, A., Mares, J. and R$\acute{\rm e}$vai, J. (2007) 
                    $\bar{K}NN$ quasibound state and the $\bar{K}N$ 
                    interaction: Coupled-channels Faddeev calculations of the 
                    $\bar{K}NN-\pi\Sigma N$ system.
                    Phys. Rev. {\bf C 76},  044004.
\bibitem{Nelson} Kaplan, D.B. and Nelson, A.E. (1986) 
                    Strange goings on in dense nucleonic matter.
                    Phys. Lett. B {\bf 175}, 57-63. 
\bibitem{Brown} Brown, G.E., Lee, C.H., Rho, M. and Thorsson, V. (1994) 
                    From kaon-nuclear interactions to kaon condensation.
                    Nucl. Phys. A {\bf 567}, 937-956.
\bibitem{Brown-Bethe} Brown, G.E. and Bethe,  H.A. (1994) 
                    A scenario for a large number of low-mass black holes in the galaxy. 
                    Astrophys. J. {\bf 423}, 659. 
\bibitem{Kienle07} Kienle, P. (2007) 
		Experimental studies of antikaon mediated nuclear bound systems.
		Int. J. Mod. Phys. {\bf A22}, 365-373
\bibitem{Zmeskal09} Zmeskal, J., B\"uler, P., Cargnelli, M., Ishiwatari, T., Kienle, P., Marton, J., Suzuki, K., 		Widmann, E. (2009)
 		Double-strangeness production with antiprotons.
                   Hyperfine Interactions {\bf 194}, 249-254.
\bibitem{Hassanvand11} Hassanvand, M., Akaishi, Y. and Yamazaki, T. (2011)
		Formation of $K^-K^- pp$ in high-energy proton-proton collisions.
		 Phys. Rev. C (submitted).
\bibitem{Akaishi86} Akaishi, Y. (1986) 
                    Few-body system in realistic interaction.
                    Int. Rev. Nucl. Phys. {\bf 4}, 259-393.
\bibitem{Kanada} Kanada-En'yo, Y., Jido, D. (2008) 
                    $\bar{K}\bar{K}N$ molecular state in a three-body 
                    calculation. Phys. Rev. C {\bf78}, 025212-10.
\bibitem{KK} Beane, S.R., Luu, T.C., Orginos, K., Parreno, A., Savage, M.J., Torok, A., and Walker-Loud, A. (2008) $K^+K^+$ scattering length from lattice QCD.
                    Phys. Rev. D {\bf 77}, 094507-11.
\bibitem{Nishijima:06} Nishijima, K. (2006) Conception of generations. 
                    Prog. Theor. Phys. Suppl. {\bf 164}, 28-37.
\bibitem{BM} Braun-Munzinger, P., Heppe, I.,  Stachel, J. (1999) 
                    Chemical equilibration in Pb+Pb collisions at the SPS.
                    Phys. Lett. B {\bf 465}, 15-20.
\end{thebibliography}
\end{document}